\documentclass[12pt]{article}

\usepackage[colorlinks,linkcolor=blue,citecolor=blue,urlcolor=blue,bookmarks,bookmarksnumbered]{hyperref}
\usepackage{amsmath,amssymb,amsfonts,mathrsfs}
\usepackage[paper=letterpaper,margin=1.0in]{geometry}
\usepackage{graphicx,xcolor}
\usepackage{cite}
\RequirePackage{tikz}
 \usetikzlibrary{arrows}
 \def\TikZ#1{\begin{tikzpicture}#1\end{tikzpicture}}

\def\join{\mathop{\texttt{\#}}\limits}
\def\AdS{\text{AdS}}
\def\SL{\mathop{\textrm{SL}}}
\def\IC{\mathbb{C}}
\def\sC{\mathscr{C}}
\let\d=\delta
\def\rd{\mathrm{d}}
\def\dS{\text{dS}}
\let\e=\epsilon
\let\F=\Phi
\let\L=\Lambda
\def\Lb{\Lambda_b}
\def\sO{\mathscr{O}}
\let\w=\omega
\let\p=\pi
\def\IP{\mathbb{P}}
\let\q=\theta
\def\IR{\mathbb{R}}
\let\s=\sigma
\let\sss=\scriptscriptstyle
\def\fS{\mathfrak{S}}
\let\t=\tau
\let\vd=\partial
\def\fX{\mathfrak{X}}
\def\sX{\mathscr{X}}
\def\sY{\mathscr{Y}}
\def\sZ{\mathscr{Z}}
\def\ZZ{\mathbb{Z}}
\def\define{\mathrel{\buildrel{\rm def}\over=}}
\let\Tw=\widetilde
\def\vev#1{\langle #1 \rangle}
\def\frc#1#2{{\textstyle{#1\over#2}}}
\parskip\medskipamount
\begin{document}
\setcounter{page}{0}
\thispagestyle{empty}
\begin{center}

\vskip 1.0cm
\centerline{\Large \bf Stringy Bubbles Solve de~Sitter Troubles}
\bigskip
\renewcommand{\thefootnote}{\fnsymbol{footnote}}
\centerline{{\bf
Per Berglund${}^{1}$\footnote{\tt per.berglund@unh.edu},
Tristan H\"{u}bsch${}^{2}$\footnote{\tt thubsch@howard.edu}
and
Djordje Minic${}^{3}$\footnote{\tt dminic@vt.edu}
}}
\bigskip\footnotesize
{\it
${}^1$Department of Physics and Astronomy, University of New Hampshire, Durham, NH 03824, U.S.A. \\
${}^2$Department of Physics and Astronomy, Howard University, Washington, D.C.  20059, U.S.A. \\
${}^3$Department  of Physics, Virginia Tech, Blacksburg, VA 24061, U.S.A. \\
${}$ \\
}
\end{center}
\vskip 1.0cm
\begin{abstract}
Finding four-dimensional de~Sitter spacetime solutions in string theory has been a vexing quest ever since the discovery of the accelerating expansion of the universe. 
Building on a recent analysis of bubble-nucleation in the decay of (false-vacuum) AdS 
backgrounds where the interfacing bubbles themselves exhibit a de~Sitter geometry we show that this resonates strongly with a stringy cosmic brane  construction that  naturally provides for  an exponential mass-hierarchy, and localization of both gravity and matter, in addition to an exponentially suppressed positive cosmological constant. Finally, we argue that these scenarios can be realized in terms of a generalization of a small resolution of a conifold singularity in the context of a (Lorentzian) Calabi-Yau 5-fold, where the isolated (Lorentzian) two complex dimensional Fano variety is a four-dimensional de~Sitter spacetime.
\end{abstract}

\bigskip

\begingroup
\baselineskip=9pt
 \parskip=\smallskipamount
\tableofcontents
\endgroup

\thispagestyle{empty}

\renewcommand{\thefootnote}{\arabic{footnote}}
\vfill


\section{Introduction}
\label{s:IRS}
For almost a quarter of a century, a specter has been haunting string theory: the accelerated expansion of our universe implies an asymptotically and approximately de~Sitter (dS) geometry with a small but positive cosmological constant~\cite{Riess:1998cb, Perlmutter:1998np}. Constructing  solutions in string theory with these features has been argued to be notoriously hard if not impossible: see~\cite{Danielsson:2018ztv,Cicoli:2018kdo,BHM:2021review} for recent comprehensive reviews. 
However,  recent work~\cite{Banerjee:2018qey,Banerjee:2019fzz} and~\cite{Bento:2021nbb}, as well as~\cite{Blaback:2019zig},  turn out to correspond naturally with a stringy cosmic dS-brane toy-model~\cite{rBHM1,rBHM4,rBHM5,rBHM7,rBHM10} which we furthermore connect with a generalization of the proposal that allows for more general spacetime varying string vacua~\cite{rHitch}.

 The main ideas are the following:
 ({\small\bf1})~Both the \AdS-decaying~\cite{Banerjee:2018qey,Banerjee:2019fzz} 
  and the so called ``axilaton''~\cite{rBHM1,rBHM4,rBHM5,rBHM7,rBHM10} scenarios are {\em\/near\/} a singularity: They are ``resolutions/smoothings'' of a singular/critical model, and so are inherently non-perturbative
~\cite{Bento:2021nbb}. 
  As such, a closer examination can (and does) find special configurations in which the various competing contributions do allow for a four-dimensional (sub-)spacetime of the desired dS geometry.
 ({\small\bf2})~Just as in the \AdS-decaying scenario, the axilaton models can be made dynamical by choosing their anisotropy, $\w$, differently on the two sides of the candidate for the observed universe, $\dS^{1,3}_{z=0}$.
 ({\small\bf3})~Calabi-Yau 5-folds, $\fX_5$, provide an auxiliary Euclidean rendition of the total Lorentzian spacetime, $\sX^{1,9}$, and a natural ``home'' for such phase-transition nucleating ``bubble-worlds'': Just as in the \AdS-decaying scenario, the ``exceptional'' (real codimension-six) sub-spacetimes $W^{1,3}$ naturally have positive curvature, while their complement and local neighborhood in $\sX^{1,9}$ is naturally hyperbolic.

The key effects of considering the theory near a singularity are discussed in Section~\ref{s:NS}, especially focusing on the conifold in~\ref{s:wdC}, which leads us to  review the \AdS-decaying scenario in~\ref{s:dSB}. In Section~\ref{s:axilaton} we recall the salient features of the axilaton toy models and highlight the similarities in the overall spacetime geometry with the AdS-decaying solutions. This in turn lets us modify the former so as to afford a dynamical scenario akin to the latter.
 The global geometry of the ten-dimensional spacetime is reviewed in Section~\ref{s:CY5}, where we present evidence that the most generic of spacetime geometries in string theory necessarily include exceptional four-dimensional sub-spacetimes of positive curvature, the simplest of which being $\dS^{1,3}$.
  Finally, section~\ref{s:Coda} summarizes our key points and conclusions, while Appendix~\ref{s:ED} collects a few additional details about (spacetime) defects and their role in understanding dS solutions in string theory.
 
\section{Nearly Singular Spacetimes}
\label{s:NS}
The early discovery that certain singular features in spacetime are innocuous to string dynamics~\cite{Dixon:1985jw,Dixon:1986jc} shows that the choice of available geometries is, in this respect, considerably more general than in conventional quantum field theory. A large class of such singular geometries affords topology change~\cite{rGHC,Green:1988uw,Candelas:1989ug,Partouche:2000uq}, in a milder sense also~\cite{Aspinwall:1993nu,Aspinwall:1994zd}, and so involve some form of a phase transition. This tends to harbor radical effects both in physics and in geometry: massless black holes~\cite{Strominger:1995cz}, exoflops~\cite{Aspinwall:1993nu,Hubsch:2002st}, $D$-branes~\cite{Polchinski:1996fm}, and orientifolds~\cite{Sagnotti:1987tw,Horava:1989vt,Bianchi:1990tb}, and often involve strong coupling effects.

\subsection{Warped Deformed Conifolds and Alike}
\label{s:wdC}
Models that are even just {\em\/near\/} being singular then invariably involve multiple competing contributions to the effective (target-spacetime) action, some of which are often difficult to calculate. Nevertheless, suitable approximations have been carefully considered and well justified, as is the case of the warped deformations of conifold singularities~\cite{Klebanov:2000hb,DeWolfe:2002nn,Douglas:2007tu}; see also~\cite{Douglas:2009zn,Bena:2009xk}. Here, the complex structure deformation of the conifold singularity and the relative size of associated vanishing cycles, the $\overline{\text{D3}}$-brane contributions, and the metric warping all affect the target spacetime physics.  A detailed examination of such a nearly singular warped direct product compactification scenario finds a fine-tuned regime within the parameter space in which the multiple competing contributions can be shown to induce a metastable dS metric in the non-compact spacetime factor~\cite{Bento:2021nbb}; see however also the review~\cite{Bena:2017uuz} and the references therein. 
 Note that the vanishing cycles in the deformation-singularizing complex, compact Calabi-Yau 3-folds are real three-dimensional subspaces, matching in dimension the ``probing'' $\overline{\text{D3}}$-branes. Thus, this example indicates that vacua with a stringy dS spacetime could require a highly specialized set of circumstances and qualities. 

Mirror symmetry guarantees that the same type of phenomenon occurs near the conifold singularity on the K{\"a}hler (or symplectic) structure side, as is the case with very small relative sizes of so-called exceptional sets\footnote{In a standard ``blowup surgery,'' these exceptional sets have long since been familiar in physics as gravitational instantons~\cite{Eguchi:1978gw,Gibbons:1978tx}, while their higher codimension ``small resolution'' analogues show up as the so-called worldsheet instanton ``$\sO(-1,-1)$-curves'' in Calabi-Yau 3-folds~\cite{rBeast}.} in a Calabi-Yau 3-fold compactification. As with the near singularization by complex structure deformation~\cite{Bento:2021nbb}, one again expects a detailed balance of various contributing factors.
 While exact calculations on this, K{\"a}hler, side are much harder in general and not amenable to explicit probing and direct computation, such mirror-symmetric configurations are qualitatively identical. In turn, however, Calabi-Yau 3-folds are known to generically have many such exceptional sets, real two- and four-dimensional subspaces, each one of which ``resolves'' a singularity by replacing it with a complex subvariety while preserving the complex structure outside the surgery locus~\cite{rH-AG,rReidK0,rBeast}; by mirror symmetry, the (nearly) vanishing 3-cycles are then as abundant.
This implies that the pool of such nearly singular models, on both mirror-symmetry sides, is populous albeit {\em\/sub-generic,} as the individual such models involve highly specialized arrangements.

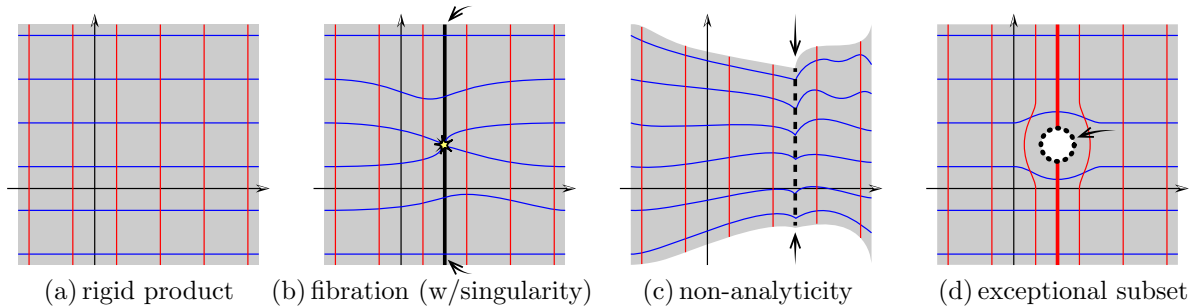
\begin{figure}[tbp]
 \begin{center}
  \begin{picture}(160,35)
   \put(0,0){\includegraphics[width=160mm]{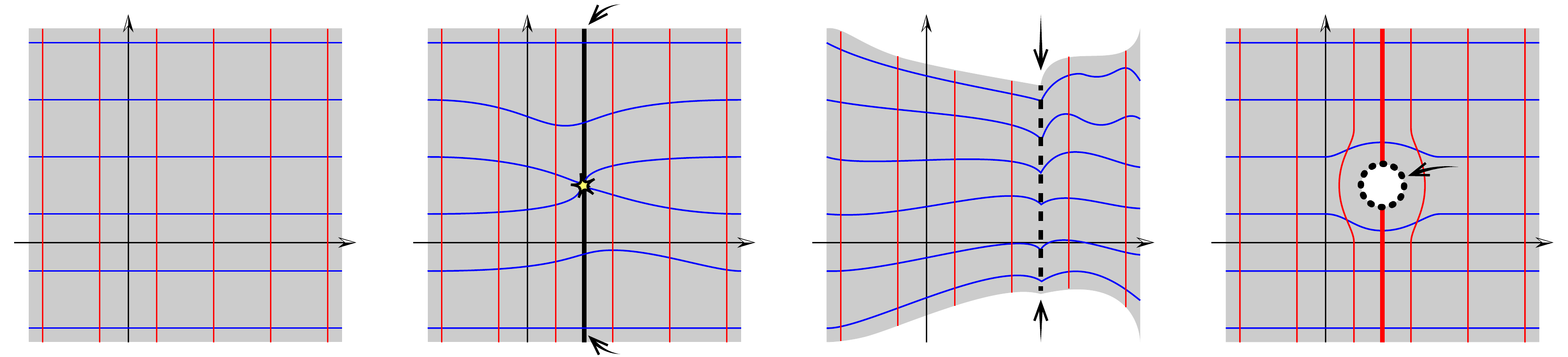}}
   \put(19,-3){\makebox[0pt][c]{\footnotesize(a)\,rigid product}}
   \put(58,-3){\makebox[0pt][c]{\footnotesize(b)\,fibration (w/singularity)}}
   \put(100,-3){\makebox[0pt][c]{\footnotesize(c)\,non-analyticity}}
   \put(142,-3){\makebox[0pt][c]{\footnotesize(d)\,exceptional subset}}
  \end{picture}
 \end{center}
 \caption{Some possible geometries in target spacetime; these simplified illustrations hint at the variations in any of the relevant structures (complex, K{\"a}hler, symplectic, supersymmetry), and may be combined in various ways}
 \label{f:4Cases}
\end{figure}

Generally, the examples discussed so far are static background configurations, wherein the observable four-dimensional spacetime\footnote{A candidate for the observable four-dimensional world with its geometry unspecified is denoted $W^{1,3}$, while $M^{1,3}$,
 $\dS^{1,3}$ and $\AdS^{1,3}$ specify Minkowski, de~Sitter and anti  de~Sitter geometries, respectively.}, $W^{1,3}$, and the extra dimensions ($\sY^6$) form a rigid direct product, as sketched at Figure~\ref{f:4Cases}\,(a), where the geometry of $\sY^6$ and of the observable four-dimensional spacetime are independent of each other.
Allowing for $\sY^6$ to vary in some observable {\em\/spatial\/} directions was shown to necessarily include singularizations of $\sY^6$ that manifest as stringy cosmic strings ('branes)~\cite{Greene:1989ya,rCYCY}, as sketched in Figure~\ref{f:4Cases}\,(b). Allowing furthermore for {\em\/non-analytic\/} variations of the geometry of some of the extra dimensions~\cite{Randall:1999ee,Randall:1999vf} warps the overall geometry and can allow for a {\em\/simultaneous\/} emergence of exponential mass-hierarchy, localization of gravity, matter, and an exponentially suppressed cosmological constant in $\dS^{1,3}$~\cite{rBHM1,rBHM4,rBHM5,rBHM7}; see the sketch in Figure~\ref{f:4Cases}\,(c). We will return to these constructions in Section~\ref{s:axilaton}.

\subsection{de~Sitter Bubble-Worlds}
\label{s:dSB}
Consider now, in turn, the recent {\em\/dynamical\/} proposal with the spacetime geometry consisting of two copies of $\AdS^{1,4}$, glued together non-analytically across a 3-brane-world, $\dS^{1,3}_{z=0}$~\cite{Banerjee:2018qey,Banerjee:2019fzz}; see also~\cite{Hawking_2000, Karch_2001,Danielsson:2021tyb}; herein, we relabel coordinates so the shell is located at $z\!=\!0$. The remaining five dimensions compactified on a suitable space of positive curvature, such as $S^5$, complete a direct product ten-dimensional spacetime, as expected in string theory.

 The hallmark novelty in the new ``\AdS-decaying'' proposal~\cite{Banerjee:2018qey,Banerjee:2019fzz} as compared with earlier literature is that the key subspace, $\dS^{1,3}_{z=0}$, emerges as a ``bubble,'' a domain wall-like interface between a metastable, false-vacuum $\AdS^{1,4}_{\rm out}$ and the nucleated, expanding true-vacuum $\AdS^{1,4}_{\rm in}$. The discontinuity at $z\!=\!0$ in the total space must be sourced (via standard matching conditions) by a suitably chosen but otherwise unspecified matter distribution, which then also provides the interfacing shell, $\dS^{1,3}_{z=0}$, with a constant tension. It is then further verified that neither bulk nor shell-localized matter invalidate the main result, that the ``shellworld,'' $\dS^{1,3}_{z=0}$, is indeed a dS  universe with the cosmological constant that can have phenomenologically acceptable values.
 As in~\cite{Randall:1999vf}, the graviton is shown to obey a Schr{\"o}dinger-like equation, which guarantees the existence of a graviton mode duly localized to the shell, $\dS^{1,3}_{z=0}$, although here this is {\em\/not\/} the mode trapped by the $\d(z)$-function well~\cite{Banerjee:2018qey,Karch_2001}. Note that the so-obtained spacetime geometry is a combination of the sketches in Figure~\ref{f:4Cases}\,(a) and Figure~\ref{f:4Cases}\,(d): $\dS^{1,3}_{z=0}$ appears as a (non-factor) subspace {\em\/within\/} the first five-dimensional spacetime factor of
 $(\AdS^{1,4}_{\rm in}\join\AdS^{1,4}_{\rm out})\times S^5$.

Studied ever since the original proposal~\cite{Coleman:1980aw} (see~\cite{Banks:2021wqu} for a recent update), such vacuum decay phenomena have been estimated to be generic in string theory~\cite{Ooguri:2016pdq,Freivogel:2016qwc}\footnote{Refs.~\cite{Ooguri:2016pdq,Freivogel:2016qwc} in fact conjecture that all de~Sitter vacua with small cosmological constant decay faster than their horizon size; the latter of these two notes the contradiction with the results of Ref.~\cite{Danielsson:2016rmq}.}, implying then the same also for nucleation, a subset of which resulting in de~Sitter bubble-worlds. These phenomena are again ``near'' phase transitions, for which no perturbative description can be expected to be complete. Notably, the analysis of Refs.~\cite{Banerjee:2018qey,Banerjee:2019fzz} focuses on the dynamically enfolding aftermath of such a phase transition, and so again pertains to nearly singular configurations. Thus, both in the claimed ubiquity and in this inherently non-perturbative nature, this is reminiscent of the string theory realization~\cite{Green:1988uw,Candelas:1989ug} of ``Reid's phantasy''~\cite{rReidK0}. \footnote{There is also a  natural connection to the more recent and rather vast cobordism generalization~\cite{McNamara:2019rup}.} In fact, as we will argue shortly, this can be made more specific in the context of spacetime varying string vacua generalizing the early work by one of the authors~\cite{rHitch}. To see why this is the case, we will however next turn to a  brief review of the axilaton toy model~\cite{rBHM5,rBHM7,rBHM10} in order to showcase the similarities with the $\AdS^{1,4}$-decay model~\cite{Banerjee:2018qey,Banerjee:2019fzz}.

\section{A Discretuum of Toy Models}
\label{s:axilaton}

Generally accommodating the overall geometry types sketched in Figure~\ref{f:4Cases}\,(a--c) and aiming for a (de~Sitter) candidate four-dimensional spacetime, the metric Ansatz has a direct sum format,
 $\rd s^2 = A^2\,\rd s^2_{\sss1,3} +B^2\,\rd s^2_{\sss6}$,
where $\rd s^2_{\sss1,3}$ and $\rd s^2_{\sss6}$ are appropriate line elements for the candidate observable spacetime, $W^{1,3}$, and its co-factor, $\sY^6$, while $A,B$ are ``warp factors'' that are most often allowed to vary over at least some directions in $\sY^6$. The particular class of toy models~\cite{rBHM1,rBHM4,rBHM5,rBHM7} further specializes to
\begin{equation}
  \rd s^2 = A^2(z)\,\rd s^2_{1,3} +\ell^2 B^2(z)(\rd z^2{+}\rd\q^2)_{\sY^2_\perp} 
  +\rd s^2_{\sY^4},
 \label{e:metric}
\end{equation}
where $\ell$ sets the length-scale in $\sY^2_\perp$, so the warp factors $A(z),B(z)$ and the coordinates $(z,\q)\in\sY^2_\perp$ are dimensionless. The observable spacetime ($\rd s^2_{1,3}$) is expected to have de~Sitter geometry, and the warp factors are chosen to depend only on the log-radial coordinate, $z$; also, $\q\simeq\q+2\p$ is the standard azimuthal angle in $\sY^2_\perp$. The last summand, $\rd s^2_{\sY^4}$, could also be warped but is for now assumed to be independent of the other six coordinates; see however below.

 This class of toy models also neglects all other matter fields, but explicitly includes the ``universal'' axion-dilaton (``axilaton'' ) field, 
 $\t\define B+ig_s^{-1}e^{-\F}$, 
in string theory, which exhibit an $\SL(2;\ZZ)$-monodromy owing to modular invariance~\cite{Polchinski:1998rq}.
Assuming then that $\t=\t(\q)$ separates variables in an evidently non-holomorphic (and so non-supersymmetric) way, and admits two 3-parameter classes of such solutions, which the $\SL(2;\ZZ)$-monodromy restricts to a {\em\/discretuum.} Notably, $\t=\t(\w\q)$, and the $\sY^2_\perp$-anisotropy, $\w$, appears in the Einstein-Friedan equation~\cite{rF79a,rF79b},
\begin{alignat}9
  R_{\mu\nu} &={\cal G}_{\t\bar\t}\,(\vd_{\mu}\t)(\vd_{\nu}\bar\t)
  \define\widetilde{T}_{\mu\nu}
  =\mathrm{diag}[\>\underbrace{0,0,0,0}_{\dS^{1,3}},\,\underset{z}{0\strut},
  \underset{\q}{(\w/2\ell)^2\strut},\,\underbrace{0,0,0,0}_{\sY^4}\>]~,
\label{e:EinStein}\\
  {\cal G}_{\t \bar\t} &=\frac{-1~~}{(\t{-}\bar\t)^2} ={\cal G}_{\bar\t\t},\quad
  {\cal G}_{\t\t}=0={\cal G}_{\bar\t\,\bar\t}.
\end{alignat}
and sources the warped geometry~\eqref{e:metric}~\cite{rBHM5,rBHM7}:
\begin{subequations}
 \label{e:newAB}
\begin{alignat}9
    A(z) &= Z_\pm(z) \Big(1-\frac{\w^2 z_0^2}{160} Z_\pm(z)^2 +\dots\Big),
  \label{e:newA}\\
   B(z) &= \frac{1}{\ell z_0\sqrt{\Lb}}
            \Big(1-\frac{\w^2z_0^2}{40} Z_\pm(z)^2 +\dots\Big)~,
  \label{e:newB}\\
\intertext{with the harmonic functions}
   Z_\pm(z) &\define 1\pm|z|/z_0,\quad z_0>0,
  \label{e:newB}
\intertext{and the cosmological constant}
   \Lb&=\frac{\omega^2 - (2\xi/3z_0)A^2(0)}{48\,\ell^2},
  \label{e:Lb}
\end{alignat}
\end{subequations}
where $0\leqslant\xi\leqslant12$ counts the source-branes\footnote{The stringy cosmic string-like~\cite{Shapere:1989kp} limit includes a total of $12{+}|\xi|$ supersymmetric 7-branes.}~\cite{rBHM2}. In~\cite{rBHM5,rBHM7,rBHM10}, the maximal, $\xi\!=\!12$, was used for simplicty. Thus, the $\sY^2_\perp$-anisotropy of the axilaton, $\t(\w\q)$, drives the cosmological constant~\eqref{e:Lb}. Furthermore, the inequalities $\w^2\!\geqslant\!(2\xi/3z_0)A^2(0)$ and $\Lb\!\geqslant\!0$ are saturated only in the supersymmetric limit, where $\w,(\xi/z_0),\Lb\!\to\!0$~\cite{rBHM5,rBHM7}. Finally, the length-scale $\ell$ emerges in~\eqref{e:EinStein} via dimensional transmutation (akin to $\L_{\rm QCD}$), as this equation is the condition (the vanishing of the worldsheet QFT beta-function, to the lowest order) for the target-spacetime metric to not renormalize.

Notably, the metric~\eqref{e:metric}--\eqref{e:newAB} is continuous at $z\to\pm z_0$ with both sign-choices in~\eqref{e:newA}, where $A(z)$ with $Z_-(z)$ merely ``bounces'' as does the flat-space line element in spherical coordinates at the origin. In stark contrast, assuming $\rd s^2_{1,3}$ to be the Minkowski line element forces the solution with $Z_-(z)$ to exhibit a naked singularity at $|z|\!=\!z_0$~\cite{rBHM1,rBHM2}, beyond which the metric becomes complex. Thereby, the de~Sitter metric~\eqref{e:metric} explicitly desingularizes the total spacetime in this class of toy models~\cite{rBHM5}. This is corroborated by verifying that the standard curvature invariants for~\eqref{e:metric}--\eqref{e:newAB} are all finite, whereas the Kretschmann (Riemann-squared) invariant diverges at $z_0$ in the Minkowski solution. Finally,~\eqref{e:newB} implies that the cosmological constant effectively specifies the ``distance'' (in the parameter space) to the so-resolved singularity.

In this sense, this class of toy models also describe nearly singular spacetimes, much as those studied in Ref.~\cite{Bento:2021nbb,Douglas:2009zn,Bena:2009xk,Bena:2017uuz} --- except that the background configurations are here restricted to a discretuum by the $\SL(2;\ZZ)$-monodromy. Indeed, the axilaton, $\t(\w\q)$, varies over $\sY^2_\perp$ including values that oscillate about $g_se^{\F}=O(1)$, even a few orders of magnitude! For the whole model, this not only implies strong coupling, but is (owing to the $\SL(2;\ZZ)$-monodromy of $\t(\w\q)$) also discontinuous along a branch-cut, e.g., along $\q\!=\!\pm\p$. Across this branch-cut, the model non-perturbatively patches a stringy weak-coupling regime with a reciprocally strong-coupling regime --- akin to ``$S$-folds''~\cite{rHIS-nGeoCY}. Nevertheless, the effective observable string coupling,
 $\vev{g_se^{\F}}$, remains perturbatively small in the sub-spacetime
 $\dS^{1,3}_{z=0}$, preserving the impression of a weakly coupled effective field theory within this toy-model candidate for the observable universe~\cite{rBHM7,rBHM10}.

For future reference, we note that the effective Planck mass-scale, $M_4$, in $\dS^{1,3}$ is exponentially boosted~\cite{rBHM4,rBHM10},
\begin{equation}
  M_4\!^2=M_6\!^4\,\ell^2\,z_0\!^{5/8}\,\xi^{-3/8}\,e^{+\xi z_0}\,
  2\p\Gamma\!_\pm\big(\frc38;\xi z_0\big),
 \label{e:M4M6}
\end{equation}
as compared with the six-dimensional Planck mass-scale resulting after compactifying on $\sY^4$.
Here, $\Gamma\!_-$ is the ``lower'' incomplete Gamma function, and $\Gamma\!_+$ its ``upper'' complement, and
 $0\!\leqslant\!2\p\Gamma\!_\pm\big(\frc38;\xi z_0\big)\!\leqslant\!2\p\Gamma\!_\pm\big(\frc38\big)\!\approx\!14.89$. This in turn exponentially suppresses the cosmological constant,
 $\Lb\propto z_0\!^{-5/4}\,\xi^{3/4}\,e^{-2\xi z_0}M_4\!^2/\ell^2$.
 As is by now standard~\cite{Randall:1999vf}, the graviton modes are governed by an effective potential that here depends on the source-counting parameter, $\xi$, that also appears in~\eqref{e:M4M6}~\cite{rBHM4}. This potential also has a superposed $\d(z)$-function well induced from the $|z|$-dependence of the metric~\eqref{e:newA}, which guarantees the existence of a $\dS^{1,3}_{z=0}$-localized graviton mode. A more detailed computation shows that the Newton potential is corrected by $O(\ell^3/r^3)$ terms~\cite{rBHM5}. This is suppressed as compared to the na{\"\i}ve dimensional analysis estimate, $O(\ell^2/r^2)$, based on $\dim(\sY^2_\perp)\!=\!2$, and indicates that the effect depends on the (hyperbolic) curvature of $\sY^2_\perp$ near the mid-radius, $z\!=\!0$.
 Also, this non-analyticity induces a $\d(z)$-contribution in the Ricci tensor, which must be matched by matter localized to the exceptional spacetime factor $\dS^{1,3}_{z=0}$ by the same $\d(z)$-function well in the effective potential. This populates the candidate observable spacetime with universally $\d(z)$-localized modes of any matter added to this toy model.

\begin{figure}[tbp]
 \begin{center}
  \begin{picture}(130,55)(-5,0)
   \put(-20,20){\includegraphics[height=30mm]{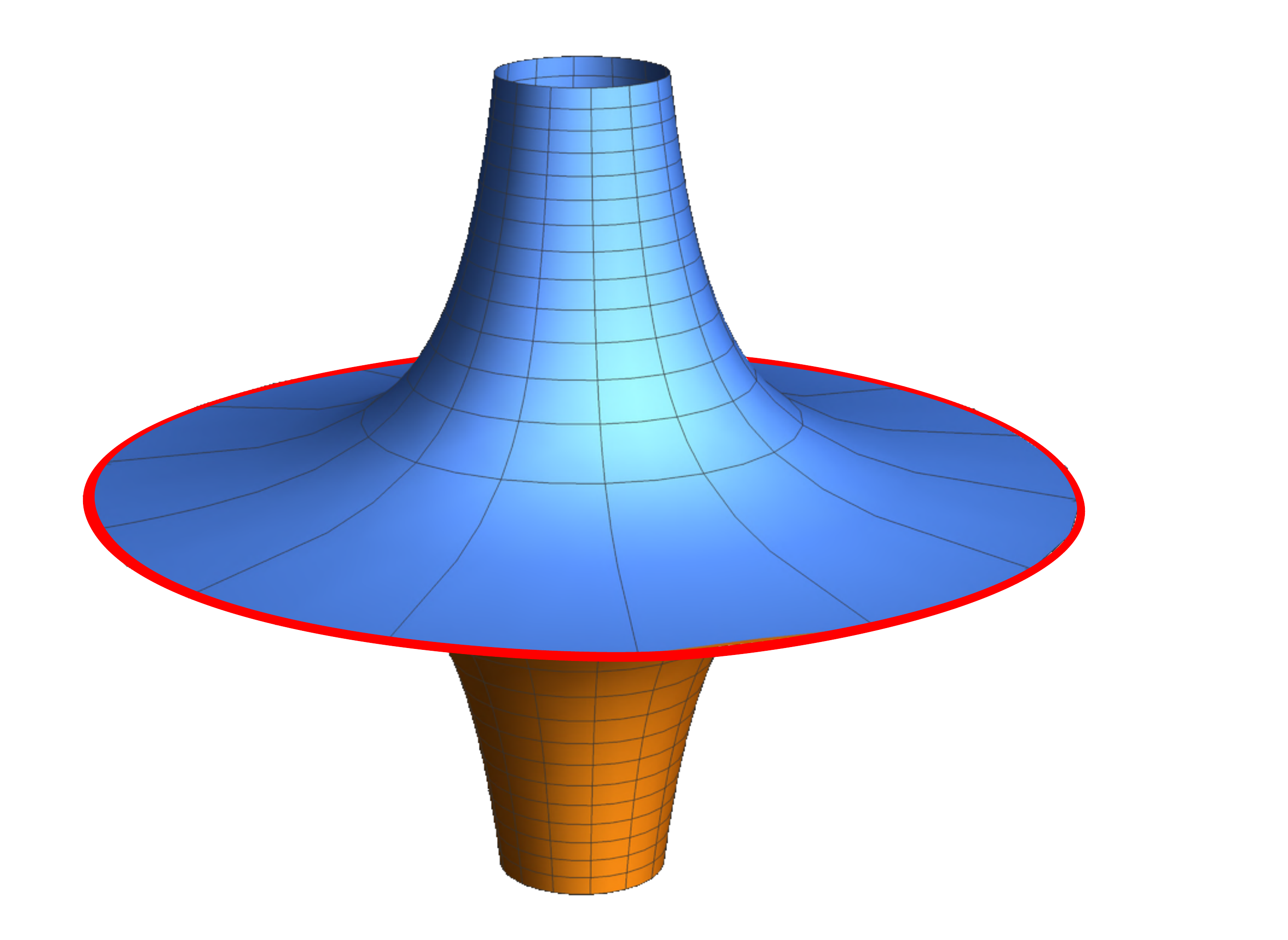}}
   \put(7,0){\includegraphics[height=50mm]{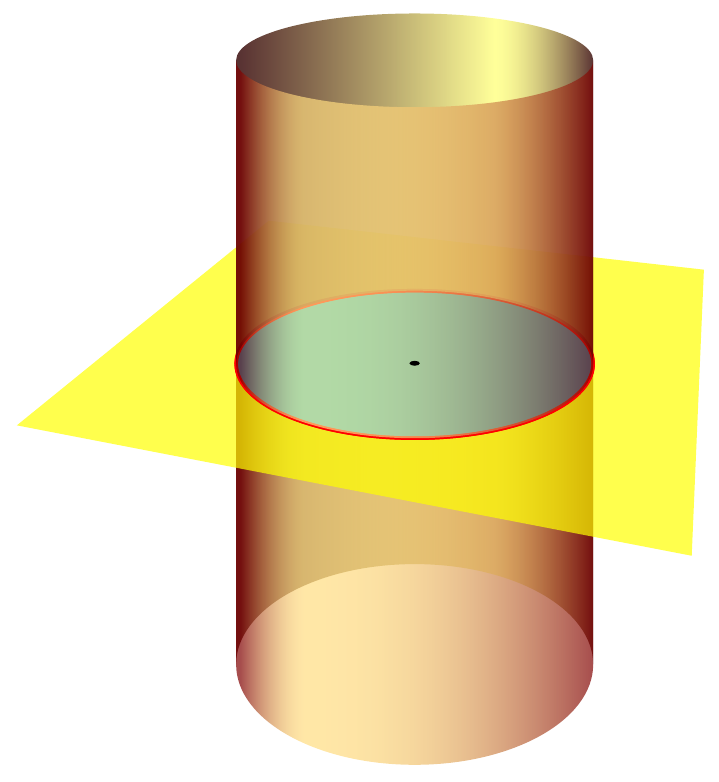}}
   \put(57,0){\includegraphics[height=50mm]{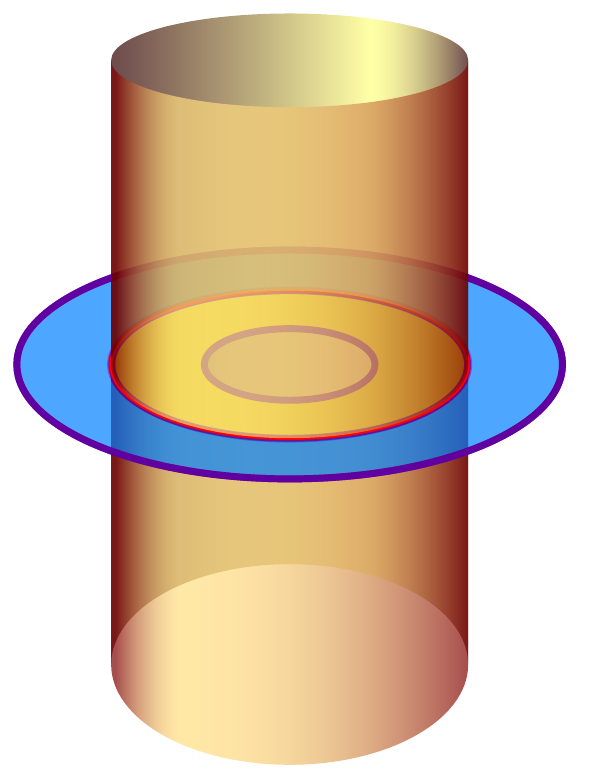}}
   \put(93,23){\includegraphics[height=25mm]{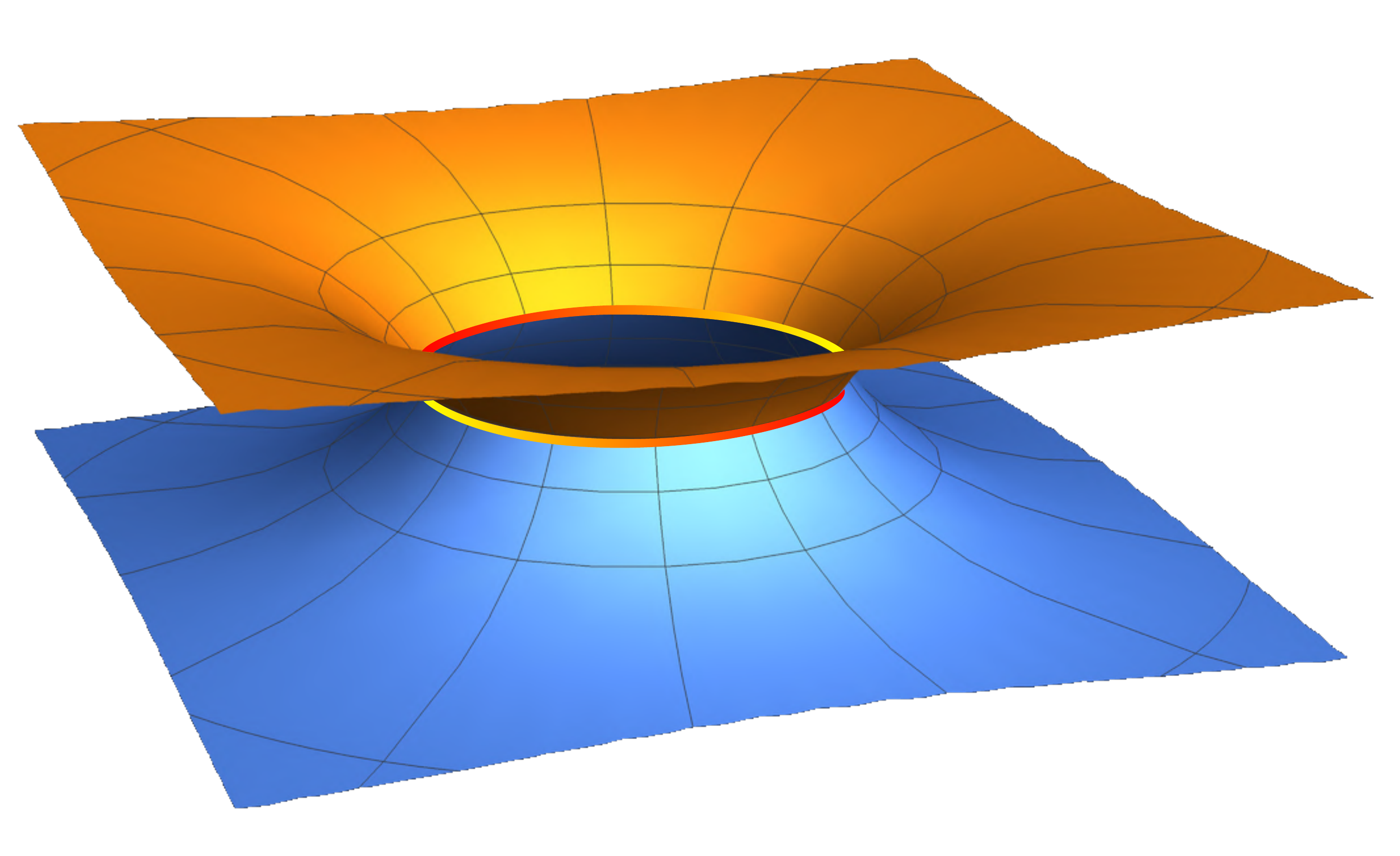}}
   \put(50,20){\TikZ{\path[use as bounding box](0,0);
              \draw[brown!75!black,ultra thick,<->](-.3,1.5)--++(0,1);
              \draw[brown!75!black,ultra thick,<->](1.2,1.5)--++(0,1);
              \path[brown!75!black](.45,2)node{\Large$W^{1,3}_{z=0}$};
              \draw[orange,thick,->](-5,.5)to[out=-60,in=-90]++(1.5,-.4);
              \draw[blue,thick,->](5,.5)to[out=-120,in=-30]++(-1.5,-.3);
              \draw[blue,thick,->](-5,2.5)to[out=30,in=90]++(3.5,-.1);
              \draw[orange,thick,->](5,2.2)to[out=120,in=90]++(-2.5,.2);
              \path(.45,.25)node{\Large$\sY^2_\perp$};
              \path(.45,-.25)node{(horizontally)};
              \path(-5,-1)node{\shortstack{$\sY^2_\perp$ with\\[2pt]
                                           $Z_+(z)=1+|z|/z_0$}};
              \path(6,-1)node{\shortstack{$\sY^2_\perp$ with\\[2pt]
                                           $Z_-(z)=1-|z|/z_0$}};
              }}
  \end{picture}
 \end{center}
 \caption{A depiction of the $W^{1,3}_z\rtimes\sY^2_\perp$ fibration with $Z_+(z)$
  (left) and $W^{1,3}_z\rtimes\sY^2_\perp$ with $Z_-(z)$ (right). Far left and far right: the proper distance plotted vertically in $\sY^2_\perp$, indicating the radial dependence of the circumference --- which is obscured in the two central depictions. Only the mid-radius fiber, $W^{1,3}_{z=0}$, of the fibration $W^{1,3}_z\rtimes\sY^2_\perp$ is depicted as a vertical cylinder, $W^{1,3}_{z=0}\times S^1$; it defines the ``inside'' and ``outside' part of the six-dimensional spacetime'; see~\eqref{e:Geometries}, below.}
 \label{f:WxY}
\end{figure}
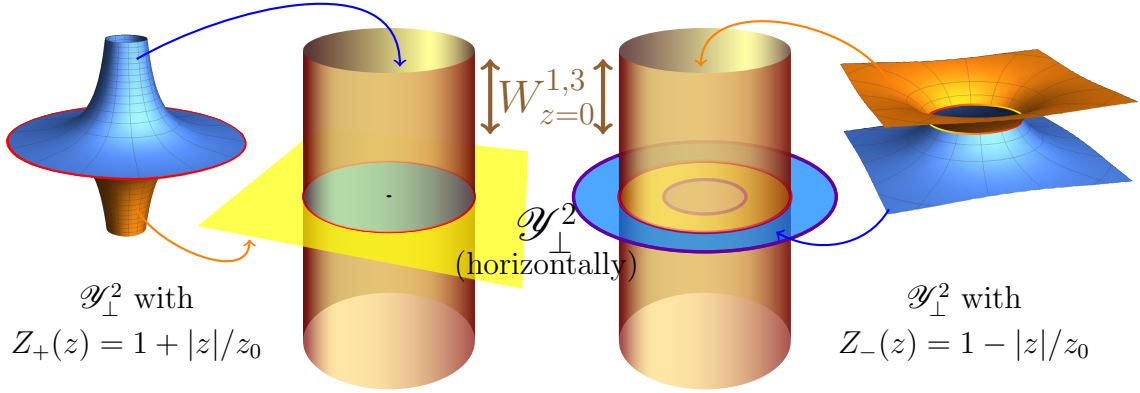

With $Z_+(z)$ in~\eqref{e:newA}, the ``transverse'' annular region, 
 $\sY^2_\perp\approx(\sZ\!\times\!S^1)$, has $\sZ_+\!=\!(-\infty,\infty)$ so that
 $\sY^2_\perp\approx\IC^*$ is non-compact, as depicted in Figure~\ref{f:WxY}\,(left).
In turn, with $Z_-(z)$ in~\eqref{e:newA},
 $\sZ_-\!=\![-z_0,z_0]$ for 
  so that $\sY^2_\perp$ includes its two circular boundaries; see Figure~\ref{f:WxY}\,(left).
In both of these solutions, the proper (log-radial) length of $\sZ$ is infinite, although the surface area of $\sY^2_\perp$ is finite.
 The 2-point compactification/completion\footnote{To be precise: for $Z_-(z)$ in~\eqref{e:newA}, each of the two circular boundaries of $\sY^2_\perp$ shrinks to a point as $z_0\to\infty$, thus rendering $\sZ_-$ compact. For $Z_-(z)$ in~\eqref{e:newA}, $\sZ\!\approx\!\IC^*$ is non-compact and two points must be added to compactify $\sZ_+\to\IP^1$.} $\sY^2_\perp\!\hookrightarrow\!S^2\!=\!\IP^1$ then allows us to regard $\sY^2_\perp$ as a 2-point {\em\/puncturing\/} of $\IP^1$. Fibering the remaining factor, $\sY^4$ (e.g., K3), over $\sY^2_\perp$ shows that the unobservable space, $\sY^2_\perp\!\ltimes\!\sY^4$ in the axilaton models~\cite{rBHM5,rBHM7,rBHM10} is also obtainable as a 2-fibre puncturing of K3-fibrations over $\IP^1$ --- the latter of which are well understood Calabi-Yau 3-folds, as they are used in static, Calabi-Yau compactifications to Minkowski spacetime~\cite{rBeast}\footnote{We will return to this non-trivial K3-fibration in Section~\ref{s:CY5}.}.
The 2-fibre compactification/completion process 
\begin{equation}
  \dS^{1,3}_z\!\rtimes\!\left(\sZ\!\times\!S^1\right)\ltimes\text{K3}
  ~\xrightarrow[~z_0\to\infty~]{~\w\to 0~}~
  M^{1,3}\!\times\!\left(\sY^6 = 
  (\IP^1\ltimes\text{K3})\right)
 \label{e:pDefo}
\end{equation}
may be understood as the right-hand side depicting a static, 
Minkowski/CY geometry being identified as the (double, supersymmetry-restoring) limit, $\w,1/z_0\to0$, of the axilaton geometry on the left-hand side. Intermediate in this process is the ``decompactification'' of the Minkowski/CY geometry if $\sY^2_\perp\!\approx\!\IC^*$ is a non-compact, unbounded cylinder, or a cylinder with boundaries if $\sY^2_\perp\!\approx\![-z_0,z_0]\!\times\!S^1$.
 In this sense, the geometry given on the far left-hand side in~\eqref{e:pDefo} is the (supersymmetry-breaking, non-holomorphic) deformation of a punctured version of the geometry depicted on the far right-hand side. In turn, the $\dS^{1,3}_{z=0}\!\times\!S^1\!\times\!\text{K3}$ configuration in the log-radial middle, $z\!=\!0$, of the annulus $\sY^2_\perp\!=\!S^1\!\times\!\sZ$ serves as a ``cylindrical'' interface between the ``inside'' and ``outside'' regions; see Figure~\ref{f:WxY}.
  
 Although somewhat different in the technical details, the overall spacetime geometry in the axilaton toy model~\eqref{e:metric}--\eqref{e:M4M6}~\cite{rBHM5,rBHM7,rBHM10} and in the $\AdS^{1,4}$-decay model~\cite{Banerjee:2018qey,Banerjee:2019fzz} are remarkably similar: The Schr{\"o}dinger-like potential governing the graviton exhibits a discontinuity at $\dS^{1,3}_{z=0}$, which then properly localizes a graviton mode.
 Just as in the axilaton toy model, the $\AdS^{1,4}$-decay model also shows the metric within the so embedded $\dS^{1,3}_{z=0}$ to be of the de~Sitter type, with the cosmological constant in the correct ballpark. The respective Ricci tensors also require a balancing source by matter modes, again localized to the same locus.
 
The two spacetime geometries, shown side-by-side, with the interfacing $\dS^{1,3}_{z=0}$ shown underscoring the joining ($\join$) symbol\footnote{The product $\mathscr{A}\!\rtimes\!\mathscr{B}$ denotes that the warp-factors in the block-diagonal metric vary over $\mathscr{B}$.}:
\begin{equation}
  \Big( \AdS^{1,4}_{\rm in}\!\join_{\dS^{1,3}_{z=0}}\!\AdS^{1,5}_{\rm out} \Big)
  \times S^5
  \quad\text{vs.}\quad
  \Big( \big( \underbrace{\dS^{1,3}_z\!\rtimes\!\sY^2_{\perp,\rm in}}_{z<0} \big)
        \!\!\join_{\dS^{1,3}_{z=0}\times S^1}\!\!
        \big( \underbrace{\dS^{1,3}_z\!\rtimes\!\sY^2_{\perp,\rm out}}_{z>0} \big) \Big)
  \times\sY^4.
 \label{e:Geometries}
\end{equation}
The left-hand side scenario is rendered dynamical by choosing
 $\AdS^{1,5}_{\rm out}$ to be the false vacuum, with higher energy than the true-vacuum
 $\AdS^{1,5}_{\rm in}$.
This inequality provides the interfacing $\dS^{1,3}_{z=0}$ with tension and drives its spatial expansion. The left-hand side scenario~\eqref{e:Geometries} thus depicts a point-nucleation of the ``true'' $\AdS^{1,4}_{\rm in}$ within the ``false'' $\AdS^{1,5}_{\rm out}$, with the 
four-dimensional boundary serving as the candidate observable spacetime, $\dS^{1,3}_{z=0}$~\cite{Banerjee:2018qey,Banerjee:2019fzz}.

 The right-hand side of~\eqref{e:Geometries} axilaton configuration has been presumed so far to be static and in/out-symmetric~\cite{rBHM5,rBHM7,rBHM10} --- but this need not at all be the case. Since all physical features in this model depend on the anisotropy of the axilaton, $\t(\w\q)$, it suffices to choose $\w_{\rm in}\neq\w_{\rm out}$. The two sides of the interfacing $\dS^{1,3}_{z=0}\!\times\!S^1$ in~\eqref{e:Geometries} are
\begin{equation}
   \big(\dS^{1,3}_z\!\rtimes\!\sY^2_{\perp,\rm in}\big)_{z<0}
   \qquad\text{and}\qquad
   \big(\dS^{1,3}_z\!\rtimes\!\sY^2_{\perp,\rm out}\big)_{z>0},
\end{equation}
which are analogous to $\AdS^{1,4}_{\rm in}$ and $\AdS^{1,4}_{\rm out}$, as depicted in Figure~\ref{f:WxY}. Indeed, the log-radial $z$-dependence renders the proper distance between $z\!=\!0$ and $z_0$ infinite while keeping the surface area of $\sY^2_\perp$ finite, implying a hyperbolic geometry. This geometry is driven by the axilaton anisotropy, $\w>0$ in $\t=\t(\w\q)$, and differs from the well-known Ricci-flat geometry of the annular $\sY^2_\perp$ that one might have expected: the negative curvature of $\sY^2_\perp$ is forced by balancing the anisotropy-driven ``pressure,''
 $\Tw{T}_{\q\q}=(\w/2\ell)^2$, in~\eqref{e:EinStein}.

 With different values of the axilaton anisotropy, the regions
 $\dS^{1,3}_{z<0}\rtimes\sY^2_{\perp,\rm in}$ and
 $\dS^{1,3}_{z>0}\rtimes\sY^2_{\perp,\rm out}$ harbor different ``pressures,'' which drives the expansion of one region into the other. 
 By analogy then, the right-hand side configuration~\eqref{e:Geometries} depicts a cosmic string-sourced nucleation of an analogous phase transition with a $\dS^{1,3}_{z=0}\times S^1$-shaped ``cylindrical'' $(3{+}1{+}1)$-dimensional interface.
 However, unlike in the \AdS-decay model~\cite{Banerjee:2018qey,Banerjee:2019fzz}, the
  $\SL(2;\ZZ)$-monodromy of the axilaton models~\cite{rBHM5,rBHM7,rBHM10} restricts the anisotropy $\w$ to discrete values. This allows only discrete variations, i.e tunneling from one configuration to another.

\section{Calabi-Yau 5-Folds}
\label{s:CY5}
Let us reconsider then the overall, total spacetime in string theory, 
relying only on the most general of restrictions while keeping the models of the foregoing discussion in the back of our mind. To this end, first generalize the geometry in the axilaton model by allowing for the K3 as in~(\ref{e:pDefo}) to be fibered over the $\sZ\!\times\!S^1$, and hence leading to a K3-fibered Calabi-Yau 3-fold in the $z_0\to\infty,~\w\to 0$ limit. 
 In addition, we consider an auxiliary, ``Euclideanized'' rendition, which affords relating algebro-geometric features of this depiction to the Lorentzian features of the original. This affords an old idea~\cite{rHitch} a fresh look, and will turn out to improve this proposal.

To start, string dynamics is generally understood to require the overall target spacetime, $\sX^{1,9}$, to be Ricci-flat --- i.e., to {\em\/admit\/} a metric the Ricci tensor of which is a total derivative. There are in general stringy corrections, but the metric is generally expected to be perturbatively ``close'' to Ricci-flat. 
 This understanding follows from several related, but rather distinct vantage points:
\begin{enumerate}
 \item the inherent semi-infinite cohomology in string theory~\cite{rFrGaZu86},
 \item the free loop space as the configuration space of (closed) strings~\cite{Bowick:1990wt,Bowick:1988nj,rBowRaj87b,rBowRaj88,rBowRaj87a,rBowRaj87,rAlGoRe87,Oh:1987sq,rHHRR-sDiffS1},
 \item the worldsheet quantum field theory approach~\cite{Pilch:1987eb}.
\end{enumerate}
The orientability of closed strings~\cite{Polchinski:1998rq} (including zero-modes~\cite{Freidel:2015pka,Freidel:2017wst,Freidel:2017nhg}!) implies that the second-listed analysis should be extended to the orientation-doubled loop-space and (Super)Diff$(S^1)/S^1$. In lieu of this extension, we note that it will perforce include the results of the original loop-space approach. (Owing to the still very much developing nature of string theory, a technically precise meaning of this notion is still unclear, and the conceptual understanding itself may require some adjustments.)
 The above studies show that the free loop-space of $\sX^{1,9}$ must be Ricci-flat, which then determines the geometry of $\sX^{1,9}$ in turn~\cite{rBeast}.
 Consider then a suitable Euclidean Wick-rotation of $\sX^{1,9}$ that is, at least in Type-II theories (owing to their $N\!=\!2$ supersymmetry) guaranteed to admit both a complex structure and a K{\"a}hler metric. This assigns to the 
 ten-dimensional spacetime, $\sX^{1,9}$, a corresponding auxiliary Calabi-Yau 5-fold, $\fX_5$; various dualities extend this argument throughout other types of string vacua.
 
Just as Calabi-Yau 3-folds generically contain numerous exceptional, isolated so-called ``$\sO(-1,-1)$ curves''~\cite{rReidK0}, Calabi-Yau 5-folds generically contain numerous exceptional, isolated Fano ($c_1,R_{\mu\bar\nu}>0$) compact complex {\em\/surfaces,} $\fS_2$, i.e., real four-dimensional subspaces~\cite{rHitch}. It the Lorentzian original, (the preimage of) at least some of these isolated subspaces are soliton-like (rather than instanton-like), and so admit a metric of the $(1,3)$-signature, perhaps singular at some (sub)locus. Owing to their positive scalar curvature\footnote{Being Fano, $c_1(\fS)\!>\!0$, implies the scalar curvature invariant to be positive, $R(\fS)>0$.}, the simplest of these will then admit a de~Sitter metric and may serve as candidates for the observable spacetime.
 Recall that for all Einstein spacetimes, $R_{\mu\nu} = \L g_{\mu\nu}$ defines $1/\sqrt{\L}$ as the characteristic length scale, relating it to the curvature, and via $\int\!\sqrt{\det(g_{\mu\nu})}$ also to the volume. Wick rotation changes the overall coefficient in this tensorial equation, and so preserves the relation.
 More precisely, it suffices for $\fS_2$ to admit a metric that equals the de~Sitter metric within the past light-cone of a typical/current observer; see Appendix~\ref{s:ED} for more detail. For such exceptional sets at least, $\Lb>0$ parametrizes the relative size of such an exceptional, isolated real 4-manifold, $(\dS^{1,3}\mapsto\fS_2)\subset\sX^{1,9}$, --- very much akin to the de~Sitter desingularizing deformation~\eqref{e:metric}--\eqref{e:M4M6}.

Reid's argument for 3-folds~\cite{rReidK0} generalizes straightforwardly to higher dimensions, so Calabi-Yau 5-folds are expected to abound in such exceptional real four-dimensional subspaces, making the pool of candidates abundant.
 Being the exceptional sets of small resolutions of conifold singularities the dynamical evolution of the Lorentzian of these four-dimensional subspaces $\dS^{1,3}\mapsto\fS_2$ may well enfold akin to the $\dS^{1,3}_{z=0}$ bubble-worlds discussed above~\eqref{e:Geometries}. In the coarse classification discussed in Section~\ref{s:NS} and referring to Figure~\ref{f:4Cases}, the scenario $\dS^{1,3}_{z=0}\subset\sX^{1,9}$ that is the preimage of the ``Euclideanizing'' Wick rotation assignment
\begin{equation}
  \dS^{1,3}_{z=0}\subset\sX^{1,9}
  ~\mapsto~
 \fS_2\subset\fX_5
\end{equation}
 is depicted in Figure~\ref{f:4Cases}\,(d): the total spacetime, $\sX^{1,9}$, is in no way assumed to factorize, nor even foliate\footnote{For our present purposes, a foliation $X\!\divideontimes\!Y$ means that the total space looks locally at every point as a direct product of local portions of the two factors, $X$ and $Y$.}, and $\dS^{1,3}\mapsto\fS_2$ is an isolated, exceptional sub-spacetime. On general grounds, such non-factorizing spacetimes ought to be the generic case in string theory.

Now consider the auxiliary Calabi-Yau 5-fold, $\fX_5$, a particular exceptional Fano surface in it, $\fS_2\subset\fX_5$, and its non-compact complement, $\fX_5\smallsetminus\fS_2$, as well as their respective Lorentzian counterparts: $\sX^{1,9}$, $\dS^{1,3}$ and $\sX^{1,9}\smallsetminus\dS^{1,3}$. Note that the Ricci curvature is not additive for such triples (nor is then the scalar curvature), but we have the Tian-Yau theorem~\cite{rTY1,rTY2}:
\begin{alignat}9
  \underbrace{\text{compact Fano$_{n{+}1}$}}_{c_1>0} &~\smallsetminus~
  \underbrace{\text{compact CY$_n$-fold}}_{c_1=0}    &~=~
  \underbrace{\text{non-compact CY$_{n{+}1}$-fold}}_{c_1=0}. \label{e:F-CY=CY}
\intertext{Here, the non-compact space on the right-hand side is known to admit a suitable K{\"a}hler metric/form, $J$, and a holomorphic volume-form, $\Omega$ (which in turn implies the existence of a covariantly constant spinor, i.e., global supersymmetry), that satisfy the standard relationship $J^n=\bar\Omega\!\wedge\!\Omega$ asymptotically, near the locus where the compact Calabi-Yau $n$-fold was excised.
 It would then seem reasonable to expect that, analogously:}
  \underbrace{\text{compact CY$_{n{+}1}$}}_{c_1=0}   &~\smallsetminus~
  \underbrace{\text{compact Fano$_n$-fold}}_{c_1>0}  &~\overset?=~
  \underbrace{\text{non-compact Hyp$_{n{+}1}$-fold}}_{c_1<0},
 \label{e:CY-F=AdS}
\end{alignat}
where Hyp$_{n+1}$ denotes a ``hyperbolic'' complex $n{+}1$-fold with negative Ricci curvature, again with analogous asymptotics near the locus where the compact Fano $n$-fold was excised.
 Both of these relations are well-nigh trivial and easily seen for $n\!=\!0$:
\begin{enumerate}
 \item $\IP^1\smallsetminus\{2\,\text{pts.}\}\approx\IC^*$ (a non-compact cylinder),
  which is well known to be Ricci-flat.
 \item $T^2\smallsetminus\{\text{pt.}\}$ is a 1-handled disc, and so a hyperbolic non-compact surface.
\end{enumerate}
Furthermore, we will need the following specializing generalization
\begin{equation}
   \underbrace{\text{compact CY$_{n{+}1}$}}_{c_1=0}  ~\smallsetminus~
  \underbrace{\text{isolated comp.\ Fano$_{n-k}$-fold}}_{c_1>0}  ~\overset?=~
  \underbrace{\text{non-compact Hyp$_{n{+}1}$-fold}}_{c_1<0}.
 \label{e:CY-kF=AdS}
\end{equation}
 Locally, this result follows from the so-called adjunction theorem~\cite{rBeast}: For example, $c_1(\IP^1)\!=\!2\!>\!0$ and $c_1(\text{CY}_3)\!=\!0$ imply that the local neighborhood of $\IP^1\subset\text{CY}_3$ is any one of the rank-2 bundles $\sO_{\IP^1}(\ell{-}2,{-}\ell)$, with arbitrary $\ell\!\in\!\ZZ$. Depending on the choice of $\ell$, the curvature of this normal bundle will be negative along some fibers but positive along others.
 However, we seek an {\em\/isolated\/} $\IP^1$, which has no local holomorphic deformations, and which happens precisely when $\ell\!=\!1$. In that exceptional case, the local neighborhood of that $\IP^1\subset\text{CY}_3$ has a uniformly negative Ricci curvature. The claim~\eqref{e:CY-kF=AdS} then aims to generalize this, albeit perhaps in an appropriate average sense, throughout the complement $\text{CY}_3\!\smallsetminus\!\IP^1$ --- and more generally, throughout $\fX_5\smallsetminus\fS_2$; we are not aware of a rigorous global result either way.

The so-generalized claim~\eqref{e:CY-kF=AdS} then implies that the non-compact complex 5-fold $\fX_5\smallsetminus\fS_2$ admits a metric of negative Ricci curvature. (Again, {\em\/locally,} near an isolated $\fS_2\subset\fX_5$, this is a consequence of the adjunction theorem.)
 With suitable boundary conditions to match the excised subspace $W^{1,3}\mapsto\fS_2$, the Lorentzian counterpart, $\sX^{1,9}\smallsetminus W^{1,3}$, should then also admit a metric with negative curvature
 --- analogous to $\AdS^{1,4}_{\rm out}$ in the \AdS-decaying scenario~\eqref{e:Geometries}. However, these exceptional 4-manifolds, $W^{1,3}\subset\sX^{1,9}$, have {\em\/nothing\/} ``inside,'' much as a circle inside $\IR^3$ does not carve it into two separate parts: The complement, $\sX^{1,9}\!\smallsetminus\!W^{1,3}$ is a single-component connected space since $W^{1,3}\subset\sX^{1,9}$ has real codimension~6. (This is unlike the real codimension-1 subspace in the \AdS-decay scenario that does carve the total spacetime, $\AdS^{1,5}\!\times\!S^5$, into two disconnected parts, an ``outside'' and an ``inside.'') Having only the ``outside,''
 $\sX^{1,9}\smallsetminus W^{1,3}$, these exceptional sub-spacetimes $W^{1,3}\subset\sX^{1,9}$ are rather literally akin to ``bubbles of nothing''; see~\cite{Witten:1981gj} and also~\cite{Dibitetto:2020csn}.

In the Euclidean (and holomorphic) rendition, the standard K{\"a}hler class, $J$, (every metric in that cohomology class) of any ``bubble'' (desingularizing exceptional set) $\fS_2$ is both positive over every complex submanifold of $\fS_2$ and has a positive square (``volume''): $\int_{C\subset\fS}J>0$ and $\int_{\fS}J^2>0$ are the standard conditions on the K{\"a}hler cone.
 In the simplest case, take the bubble to be $\IP^2$. Then:
\begin{enumerate}
 \item The K{\"a}hler class of $\IP^2$ is positive over ``all complex submanifolds,''
   all of which are equivalent to the $\IP^1$ at the North Pole ``infinity.'' (This refers to the standard cell decomposition $\IP^2\approx\IC^2\cup\IP^1$.)
 \item There is therefore a Riemannian metric that differs from the above
   K{\"a}hler metric only in being null over the $\IP^1$ at the North pole,
 \item which is therefore a valid metric on $(S^4 \smallsetminus \{\text{Noth pole}\})\approx\IC^2$,
   and fails in those positivity requirements only at the North pole,
   where it vanishes --- and so is nowhere negative.
\end{enumerate}
Together, these imply that $W^{1,3}$ has positive curvature, the simplest (most symmetric) of which is $\dS^{1,3}$; see also Appendix~\ref{s:ED}.

Other Fano complex surface candidates for the bubble $\fS_2$ will contain more than one exceptional (complex) curve, each isomorphic to $\IP^1\!\approx\!S^2$~\cite{rBeast}. Their Lorentzian preimages, $W^{1,3}$, will contain more than one corresponding exceptional real two-dimensional sub-spacetime, with a mutual intersection pattern characteristic to the Euclidean (and complex) $\fS_2$. With a Lorentzian metric on $W^{1,3}$, these sub-spacetimes can serve as specific loci of interest, such as space-like horizons at infinite past/future. A further exploration of such correlations between the Lorentzian metric structure of topologically nontrivial spacetimes and the (almost) complex structure of their auxiliary Euclidean renditions, but beyond our present scope.

As it is, (almost) Fano complex surfaces are $\IP^1\times\IP^1$, or a blow-up of
 $\IP^2$ at $0,{\cdots},9$ points.
 Their Euler numbers ($\int\!c_2$) are, in the given order: 4, or $3,4,{\cdots},12$~\cite{rGHSAR,rBeast}.
 (The projective space, $\IP^2$, itself may be regarded as a blowup of $S^4$, by replacing a point in $S^4$ by the exceptional hyperplane $\IP^1\subset\IP^2$.)
 By the Todd-Chern-Hirzebuch theorem for Fano surfaces,
 $\int(c_1\!^2{+}c_2)=12$,  so that $\int\!c_1^2 = 12{-}\!\int\!c_2$,
and all characteristic classes (and their integrals) are controlled by $\int\!c_2$,  i.e., the Euler number. For completeness, we note that $p_1 = c_1\!^2{-}2c_2 = 3(4{-}c_2)$, so the {\em\/signature\/} of these (real) four-dimensional spaces is $\s = 4{-}\chi_E$.
 For all (almost) Fano surfaces except $\IP^1\times\IP^1$, the above argument shows that they admit a metric that is nowhere negative, and merely becomes null at select $1,{\cdots},10$ exceptional locations, which in such a metric become ``points''; the total space of these algebraic surfaces is then diffeomorphic to $S^4 \smallsetminus \{\text{points}\}$.
  In the ``ruled (complex) surface,'' $\IP^1\times\IP^1$, the meridians of either $\IP^1$ may be identified as ``time,'' which identifies the ``Southern'' and  ``Northern'' instance of the other $\IP^1$ as the infinite past and future horizon, respectively.
 It seems plausible to expect a Lorentzian metric to exist wherein those exceptional loci are at infinite past/future, and whereby the positive curvature of of these surfaces (just as of $S^4$) should translate into a positive cosmological constant, for the most part of $W^{1,3}\mapsto\fS_2$; see Appendix~\ref{s:ED}.

Even without computational details about the embedding $W^{1,3}\subset\sX^{1,9}$, the fact that this is (a Lorentzian preimage of) a small resolution smoothing has strong consequences:
 Considering $W^{1,3}$ from ``outside,'' as a (metrically) isolated as $W^{1,3}\subset\sX^{1,9}$ may be, at least locally near its locus within $\sX^{1,9}$, the whole spacetime $\sX^{1,9}$ must itself admit a Lorentzian metric with a {\em\/co-laminar\/} class of time coordinates. That is, any particular time-like geodesic, $\sC_t\subset W^{1,3}\subset\sX^{1,9}$, will have infinitesimally near time-like geodesics, $\sC_{t,\e}\subset(\sX^{1,9}\!\smallsetminus\!W^{1,3})$, such that $\lim_{\e\to0}\sC_{t,\e}=\sC_t\subset W^{1,3}$. (This may well become ill-defined at certain special locations of $ W^{1,3}\subset\sX^{1,9}$, such as past/future horizons.)
 That is, the Lorentzian metric of $ W^{1,3}$ must extend almost everywhere smoothly into the local neighborhood of $ W^{1,3}\subset\sX^{1,9}$ and then $W^{1,3}$ itself. Thereby, the above-implied class of ``time-sliced, static'' snapshots and associated phase transition nucleation interpretation also extend to the ten-dimensional spacetime
 $\sX^{1,9}$, at least in the local neighborhood of $ W^{1,3}\subset\sX^{1,9}$.

Finally, let us note that in this (putatively de~Sitter) ``exceptional sub-spacetime, $\dS^{1,3}_{\rm us}\subset\sX^{1,9}$,'' scenario $\sX^{1,9}$ is Ricci-flat, and so admits a standard, supersymmetry-preserving, spacetime metric. The putatively negative-curvature complement, $\sX^{1,9}\!\smallsetminus\!\dS^{1,3}_{\rm us}$, may well admit even multiple covariantly-constant spinors {\em\/away from\/} $\dS^{1,3}_{\rm us}$ but their restriction to $\dS^{1,3}_{\rm us}$ will fail to be covariantly constant {\em\/within\/} $\dS^{1,3}_{\rm us}$.
 Now, in the Euclideanized complex rendition, an isolated $\fS_2\subset\fX_5$ is metrically null in the ``bulk'' K{\"a}hler metric; it needs an $\fS_2$-localized ``correction'' to be positive everywhere on $\fX_5$; see~\cite{rGHSAR} for a lower-dimensional explicit construction to this end.
 This isolation then implies that $\sX^{1,9}$ admits a metric in which $\dS^{1,3}_{\rm us}$ is null, i.e., a point, so that for a typical ``bulk''-$\sX^{1,9}$ observer, supersymmetry is broken only at a point --- a singularity of the supersymmetry structure\footnote{Supersymmetry is in string theory largely correlated with complex structure, and as mentioned above, $(S^4 \smallsetminus \{\text{pt.}\})\approx\IC^2$ of course admits a complex structure, for which the excised point is an obstruction.}.
 
Thus, these metrically isolated, codimension-six sub-spacetimes, $\dS^{1,3}_{\rm us}\subset\sX^{1,9}$, are exceptional sets of local small resolution smoothings of the ostensibly most generic rendition of ten-dimensional string vacua; in these ``measure-zero,'' but generically abundant sub-spacetimes, supersymmetry is broken.

\section{Summary, Outlook and Conclusions}
\label{s:Coda}
In this paper, we have briefly reviewed some recent work on constructing concrete superstring models that do exhibit a sub-spacetime with a phenomenologically acceptable de~Sitter geometry and cosmological constant,~\cite{Banerjee:2018qey,Banerjee:2019fzz,Bento:2021nbb}. Notably, these indicate two separate aspects that render such work more fruitful:
 ({\small\bf1})~a careful analysis of multi-parameter and near-singular configurations,
and
 ({\small\bf2})~focusing on exceptional sub-spacetimes within larger-dimensional spacetimes with non-factoring geometry.
 These two ideas in fact naturally resonate closely with the discretuum of ``axilaton'' models~\cite{rBHM1,rBHM4,rBHM5,rBHM7,rBHM10} that may be viewed as a non-holomorphic and non-analytic deformation of the stringy cosmic string ('brane) scenario~\cite{Greene:1989ya,rCYCY}.
 
 In particular, we have shown herein that this class of models exhibits both of these hallmark characteristics, and can also be adapted so as to model the dynamical scenario of~\cite{Banerjee:2018qey,Banerjee:2019fzz}. In this variant, the model represents a stringy cosmic string-sourced nucleation in a phase transition, where the candidate observable world, $\dS^{1,3}_{z=0}$, is in the interfacing boundary.

 Inspired by these resonances, we have reexamined the global geometry of the ten-dimensional spacetime in string theory, $\sX^{1,9}$, following~\cite{rHitch} and~\cite{rBHM7}. This finds plenty of exceptional sub-spacetimes that map to exceptional sets, $\fS_2\subset\fX_5$, of small resolution desingularizations of the auxiliary Euclidean, Calabi-Yau 5-fold re-rendering of the total spacetime, $\sX^{1,9}\mapsto\fX_5$. Given their positive Ricci curvature, $c_1(\fS_2)\!>\!0$, 
 the simplest of these have a de~Sitter geometry and so can serve as candidates for the observed four-dimensional spacetime. Furthermore, the four-dimensional bubble-worlds, $\dS^{1,3}_{z=0}\mapsto\fS_2$ are metrically isolated, codimension-six sub-spacetimes, that have no ``inside,'' and their local neighborhood in $\sX^{1,9}$ has a negative Ricci curvature --- on par with
 $\AdS^{1,4}_{\rm in},\AdS^{1,4}_{\rm out}$ in~\eqref{e:Geometries} in the \AdS-decaying scenario of Section~\ref{s:dSB}.
 
 It is striking to note that the above features pertain to the generic Ricci-flat ten-dimensional stringy spacetimes, each of which in turn contains a very large number of such sub-spacetimes. That is, no special requirement or restriction on the total ten-dimensional (super)stringy spacetime were needed for the above conclusions to hold, and only the general characteristics of (super)string spacetimes and their Euclideanized Calabi-Yau counterpart have been presumed. 
 

\noindent
{\bf Acknowledgments:} 
We thank Giuseppe Dibitetto and Ivonne Zavala for their kind invitation to contribute to this special issue.
 PB would like to thank the CERN Theory Group for their hospitality over the past several years.
 TH is grateful to the Department of Mathematics, University of Maryland, College Park MD, and the Physics Department of the Faculty of Natural Sciences of the University of Novi Sad, Serbia, for the recurring hospitality and resources.
 DM is grateful to Perimeter Institute for hospitality and support.
The work of PB is supported in part by the Department of Energy 
grant DE-SC0020220.
 The work of DM is supported in part by Department of Energy 
(under DOE grant number DE-SC0020262) and the Julian Schwinger Foundation.

\appendix
\section{The Effect of Defects}
\label{s:ED}
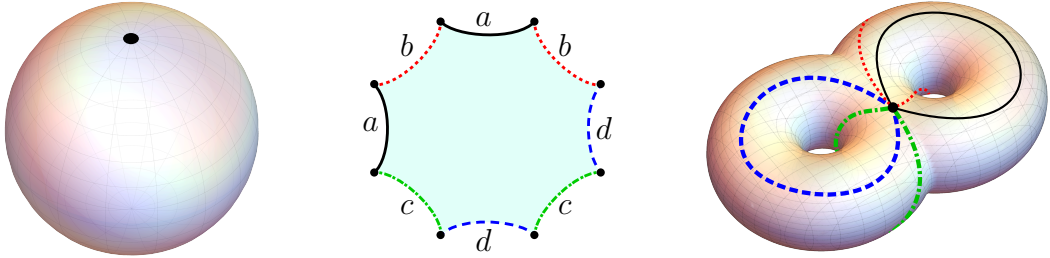
\begin{figure}[tbp]
 \begin{center}
  \begin{picture}(140,35)
   \put(0,0){\includegraphics[height=35mm]{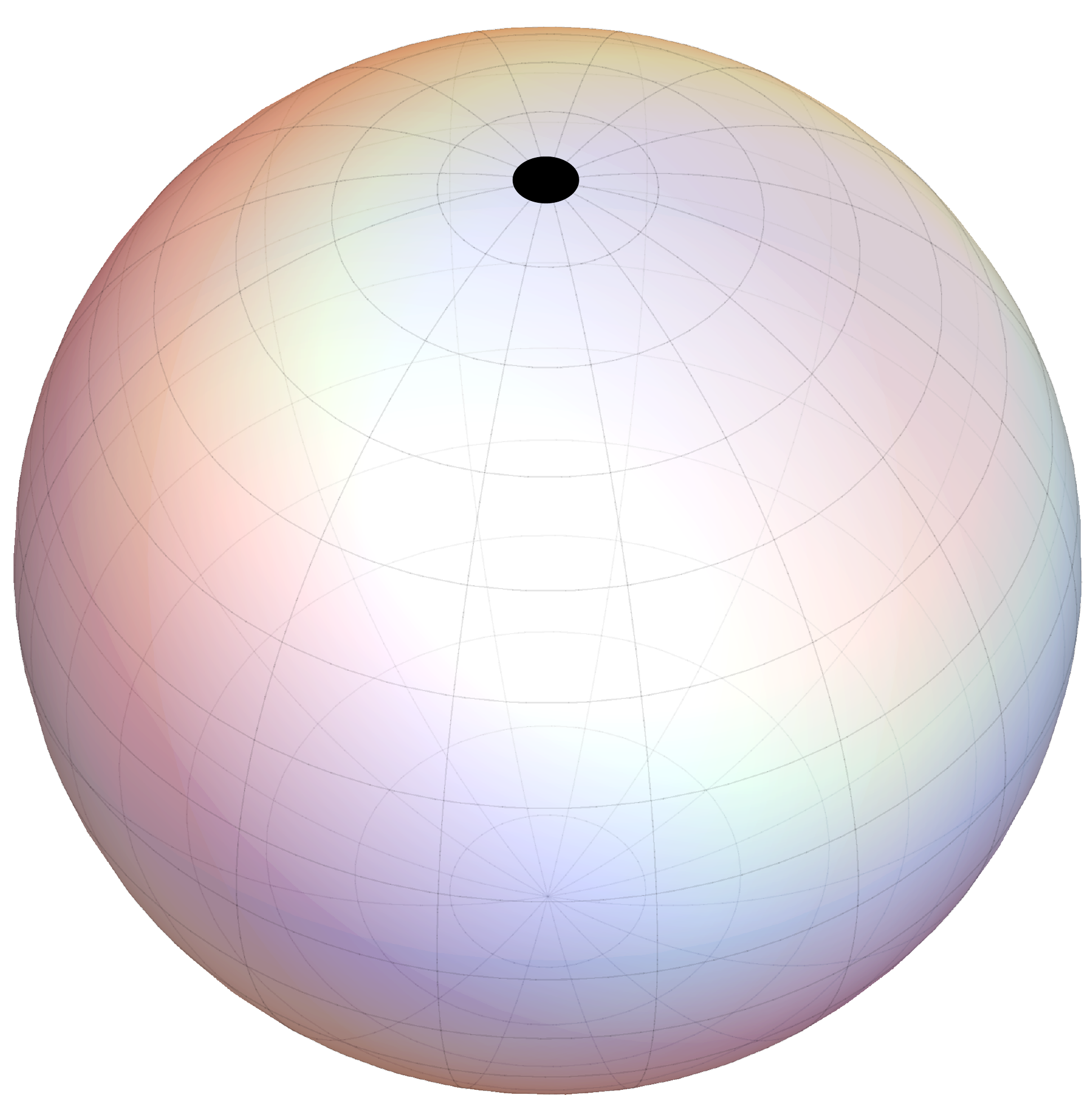}}
   \put(47,0){\includegraphics[height=35mm]{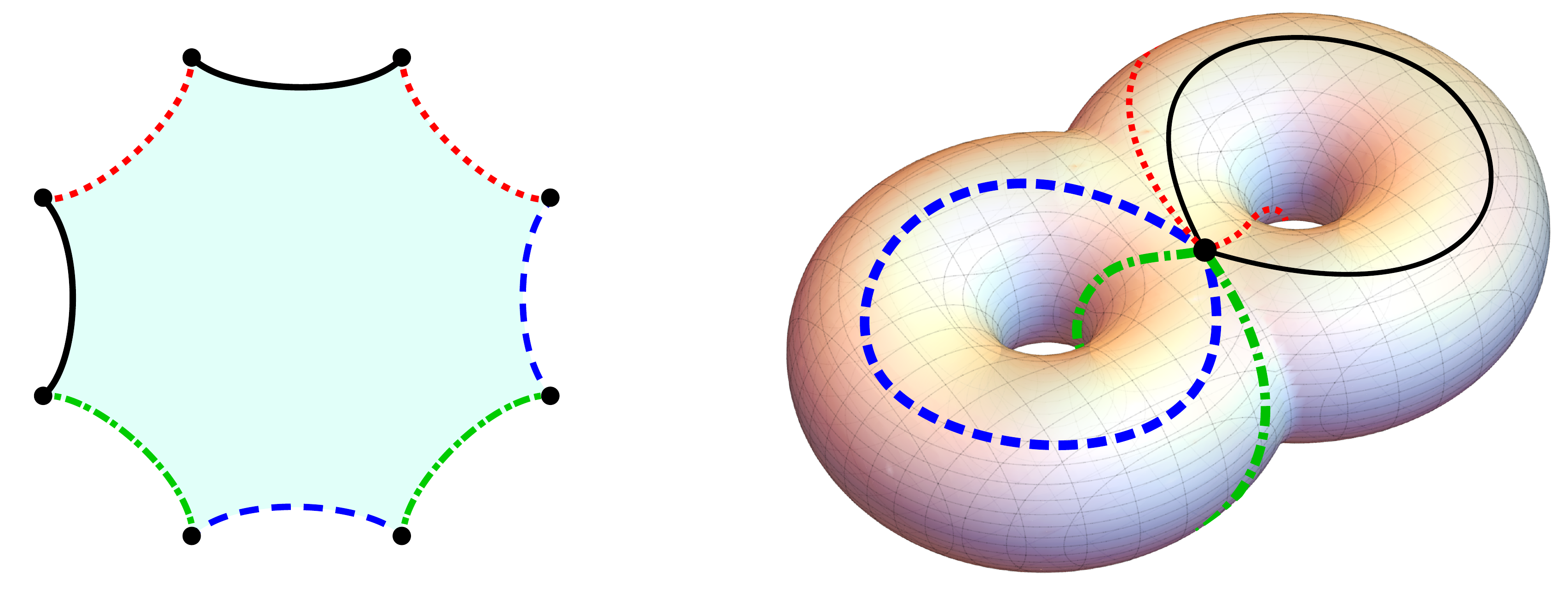}}
   \put(48,17){$a$} \put(63,31){$a$}
   \put(53,27){$b$} \put(74,27){$b$}
   \put(53,6){$c$}  \put(74,6){$c$}
   \put(63,1){$d$}  \put(79,16){$d$}
  \end{picture}
 \end{center}
 \caption{A disc (middle, suggesting a hyperbolic metric), may be compactified to a sphere (left) by identifying the entire boundary with the ``North pole.'' Identifying the boundary segments in the indicated interleaved fashion produces the genus-2 Riemann surface (right, a basis of 1-cycles identified with the indicated boundary segments).}
 \label{f:dblTorus}
\end{figure}

As a toy example, a sphere is of course positively curved and evidently admits no global metric of non-positive curvature. Excising a point (e.g., the North pole) leaves a disc, which then admits all kinds of metrics, pretty much solely depending on (controlled by) the boundary conditions. For example, identifying all the points of the boundary with a single point maps back to the sphere, identifying the disc boundary with, e.g., the North Pole; see the left-hand side of Figure~\ref{f:dblTorus}.
In turn, dividing the boundary of the disk into eight segments, which are then identified in the interleaved fashion depicted in the middle of Figure~\ref{f:dblTorus} produces a genus-2 Riemann surface depicted on the right-hand side of Figure~\ref{f:dblTorus}.
Within the differentiable category of smooth metrics, the metric on the three depicted objects need only differ in the infinitesimal neighborhood of the North pole of the sphere: Passing from the sphere (Figure~\ref{f:dblTorus}, left) to the ``double torus'' (Figure~\ref{f:dblTorus}, right) shows that the North Pole of the sphere ($H_1(S^2,\ZZ)=0$) has effectively been ``blown up''\footnote{This may be pictured as a two-step process: ({\bf1})~``open'' the point into a circular boundary, then ({\bf2})~identify segments of the boundary according to the template in Figure~\ref{f:dblTorus}, middle.} into the interleaving network of four cycles of the double torus, which span $H_1(T^2{\join}T^2,\ZZ)\approx\ZZ^4$. That is to say, the entire topological (and curvature) difference between $S^2$ and $T^2{\join}T^2$ can be concentrated into the ``$\e$-neighborhood'' of the North pole, replacing it by a 2-handled $\e$-disc:
\begin{equation}
 \vcenter{\hbox{\begin{picture}(110,15)
   \put(0,0){\includegraphics[height=15mm]{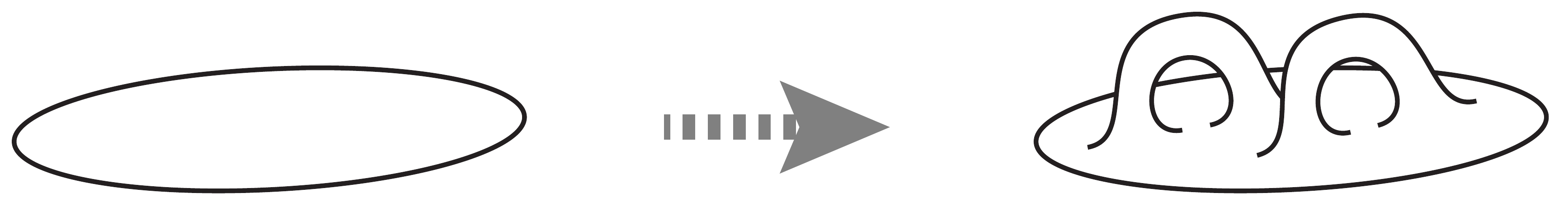}}
 \end{picture}}}
 \label{e:2handles}
\end{equation}

In the Section~\ref{s:CY5} scenario where the candidate for our own Universe is a bubble-world, $W^{1,3}$, within the Lorentzian preimage of the auxiliary Calabi-Yau 5-fold, $\sX^{1,9}\mapsto\fX_5$, we are experiencing our bubble-world with time passing along a ``meridian.'' For example, in Figure~\ref{f:dblTorus}\,(left), our bubble-world cosmic time may well be aligned with the meridians, starting at the South pole (identified as the Big Bang) and advancing towards the North Pole (identified as The End). In that case, the topological ``handles'' confined within an $\e$-neighborhood of the North pole could not have been observed by any observer, throughout almost all of the history of such a bubble-world.
 The same then applies to any topological obstruction to a de~Sitter metric. Therefore, it suffices that a candidate for our spacetime, $W^{1,3}$, admit a Lorentzian metric that is indistinguishable from the de~Sitter metric within our past light cone.
 
 Furthermore, since in such a scenario the observed matter spectrum correlates with certain specific (co)homology groups valued both in intrinsic ($T_W,T^*_W$) and in extrinsic ($N_{W\subset\sX}$) structures pertaining to our bubble-world, $W^{1,3}\subset\sX^{1,9}$, at least some of those may well pertain to topological (and (co)homological, etc.) features that are still outside of our past light cone, i.e., they could not have been observed yet. This reasoning implies it possible that, say, we have not observed a 4th generation of Standard Model particles {\em\/yet,} but may observe them sometime in the future, once the requisite topological ``handle'' in $W^{1,3}$ enters our past light cone. Since the requisite (co)homology elements are localized to those ``handles,'' they will then be the sources of such ``new'' matter. The ``time-sliced, static'' snapshots then appear to show a nucleation of a phase transition: with a 3-generation Standard Model vacuum all around, and a 4-generation vacuum nucleating and emanating from that particular ``handle.''

%

\def\rasp{\leavevmode\raise.45ex\hbox{$\rhook$}}
\providecommand{\href}[2]{#2}\begingroup\raggedright\endgroup

\end{document}